\documentclass[twocolumn,aps,pra]{revtex4-1}
\usepackage{epsfig}
\usepackage[english]{babel}
\usepackage{latexsym}
\usepackage{graphics}
\usepackage{subfigure}
\usepackage{epsfig}
\usepackage{graphicx}
\usepackage{dcolumn}
\usepackage{amsmath}
\usepackage{hyperref}
\usepackage{amssymb}
\usepackage{appendix}
\usepackage{color}
\usepackage{lineno}
\usepackage{soul}
\usepackage{booktabs}
\usepackage{longtable}
\usepackage{multirow}
\usepackage{graphics}
\usepackage{array}
\usepackage{ulem}
\begin{document}

\title{“Fish bone” resonance structure in the attosecond transient absorption spectroscopy of graphene}
\author{Fulong Dong, Jie Liu$^{*}$}

\date{\today}

\begin{abstract}
We investigate the attosecond transient absorption spectroscopy (ATAS) of graphene by numerically solving four-band density-matrix equations, which demonstrates apparent “fish bone” resonance structures. 
To gain insight into these interesting structures, we exploit a simplified model that  only considers the electrons of $\Gamma$ and $\textsc{M}$ points in the Brillouin zone. 
With the help of this model, we can analytically express the ATAS spectrum as the sum of zeroth- and first-order Bessel functions in the variables of the strength and frequency of the infrared pump field as well as the effective mass of electrons at the $\Gamma$ and $\textsc{M}$ points.
Lorentzian and Fano line shapes in the absorption spectrum are addressed.
The “fish bone” structure consists of periodic V-shaped structure that can be explained by first-order Bessel functions and its tilt angle is solely determined by the frequency of the pump laser.
The periodicity of the V-shaped structure in the “fish bone” originates from the periodic dependence of the Lorentzian and Fano line shapes of the absorption spectrum on the time delay between the pump and probe lasers.
Compared with the numerical results, our analytical theory can qualitatively or even quantitatively predict the zeroth- and first-order fringes in the “fish bone” structures of the ATAS spectrum. The gauge issues in the numerical simulations are also discussed.

\end{abstract}
\affiliation{Graduate School, China Academy of Engineering Physics, Beijing 100193, China}

\maketitle

\section{Introduction}

Recent progress in laser technology has enabled the production of an isolated pulse with a time scale down to $43$ attoseconds \cite{Ferenc,Hentschel,TGaumnitz}, which allows investigation of electron dynamics on an ultrashort time scale \cite{ALCavalieri}.
One  promising approach to investigate the subfemtosecond dynamics of electronic systems is attosecond transient absorption spectroscopy (ATAS) \cite{Eleftherios,MetteBGaarde,MengxiWu,AnneliseRBeck}, which offers an all-optical approach to reveal light-matter interactions with the high temporal resolution of the attosecond pulse and the high energy resolution characteristic of absorption spectroscopy.
ATAS has been used to study the electron dynamics of atoms and molecules \cite{PeterMKraus,RomainGeneaux,HeWang,MHoller,ZQYang,LarsBojerMadsen}, in which quantum interference \cite{ShaohaoChen}, nonresonant AC Stark shift \cite{MichaelChini} and resonant Autler-Townes splitting \cite{XiaoxiaWu,DifaYe} phenomena have been studied. More recently, attosecond time-resolved technology has been applied to bulk solids \cite{RomainGeneaux,MVolkov,MLucchini,FSchlaepfer,TOtobe,MartinSchultze,MatteoLucchini} and some  two-dimensional materials \cite{GioCistaro,ShunsukeASato}.
Since  these  materials have periodic atomic arrangements, some special resonance structures emerge in the ATAS spectrum \cite{MLucchini,ShunsukeASato}.

Graphene is a simple but specific two-dimensional material, in which there are only two carbon atoms per unit cell and the atoms are orderly arranged in a periodic hexagonal lattice. 
The unique electronic structure of graphene \cite{Castro,PRWallace} exhibits a variety of nonlinear optical processes \cite{NYoshikawa,Alonso,Dong1,Rost}.
In this work, we investigate the ATAS spectrum of graphene and address the interesting “fish bone”  resonance structure. Analytically, we can approximately express the ATAS spectrum as the sum of zeroth- and first-order Bessel functions in the variables of the strength and frequency of the infrared (IR) pump field as well as the effective mass of electrons at the $\Gamma$ and $\textsc{M}$ points. We find that the V-shaped structure of the “fish bone” can be explained by first-order Bessel functions. The periodicity of the V-shaped structure in the “fish bone” originates from the periodic dependence of the Lorentzian or Fano line shape of the absorption spectrum on the time delay between the IR pump and the attosecond X-ray probe lasers. Our analytical theory is compared with the numerical results obtained by solving four-band density-matrix equations.

This paper is organized as follows.
We describe our models and numerical results of the ATAS spectrum for graphene in Sec. \ref{s2}.
Section \ref{s3} presents the analytical formulation of the resonance structure of the ATAS spectrum.
Finally, Sec. \ref{s4} presents our conclusion.
Throughout the paper, atomic units are used if not specified.

\section{Density-matrix equations and ATAS spectrum}

\label{s2}

Graphene is a two-dimensional single layer of carbon atoms arranged in a honeycomb lattice \cite{Castro}, and it has a hexagonal lattice structure in its reciprocal space.
In this work, we consider four energy bands of graphene consisting of two core bands ($g_{1}$ and $g_{2}$), which arise from the two $1s$ orbitals of the two carbon atoms in a unit cell, and the valence ($v$) and conduction ($c$) bands, which arise from the $\pi$ orbitals orthogonal to the monolayer. 
The two core bands are degenerate and have a constant energy of $-280$ eV over the $\textbf{k}$ space.

The tight-binding Hamiltonian $H_{0}$ arising from the $\pi$ orbitals in graphene has the form
$
H_{0}=\left(\begin{array}{cc}
0 & \gamma_{0} f(\mathbf{k}) \\
\gamma_{0} f^{*}(\mathbf{k}) & 0
\end{array}\right),$
in which electrons can only hop to nearest-neighbor atoms with hopping energy $\gamma_{0}=0.1$ a.u. and $f(\textbf{k})=e^{i \texttt{k}_{x}d}+2\cos( \sqrt{3}\texttt{k}_{y}d / 2)e^{-i\texttt{k}_{x}d/2}$, with a carbon-carbon bond length of $d= 1.42$ \AA ($\approx 2.684$ a.u.).
Diagonalization of the $H_{0}$ matrix can yield energy eigenvalues, which describe the dispersion relation of the $v$ and $c$ bands $\varepsilon_{c}(\textbf{k}) = -\varepsilon_{v}(\textbf{k}) = \gamma_{0} \vert f(\textbf{k}) \vert = \gamma_{0} \sqrt{3 + 2 \cos(\sqrt{3} \texttt{k}_{y} d)+ 4\cos(3 \texttt{k}_{x} d/2)\cos(\sqrt{3} \texttt{k}_{y}d/2)}$.

%%%%%%%%%%%%%%%%%%%%%%%%%%%%%%%%%%%%%%%%%%%%%%%%%%%%%
\begin{figure}[t]
\begin{center}
{\includegraphics[width=8.5cm,height=4cm]{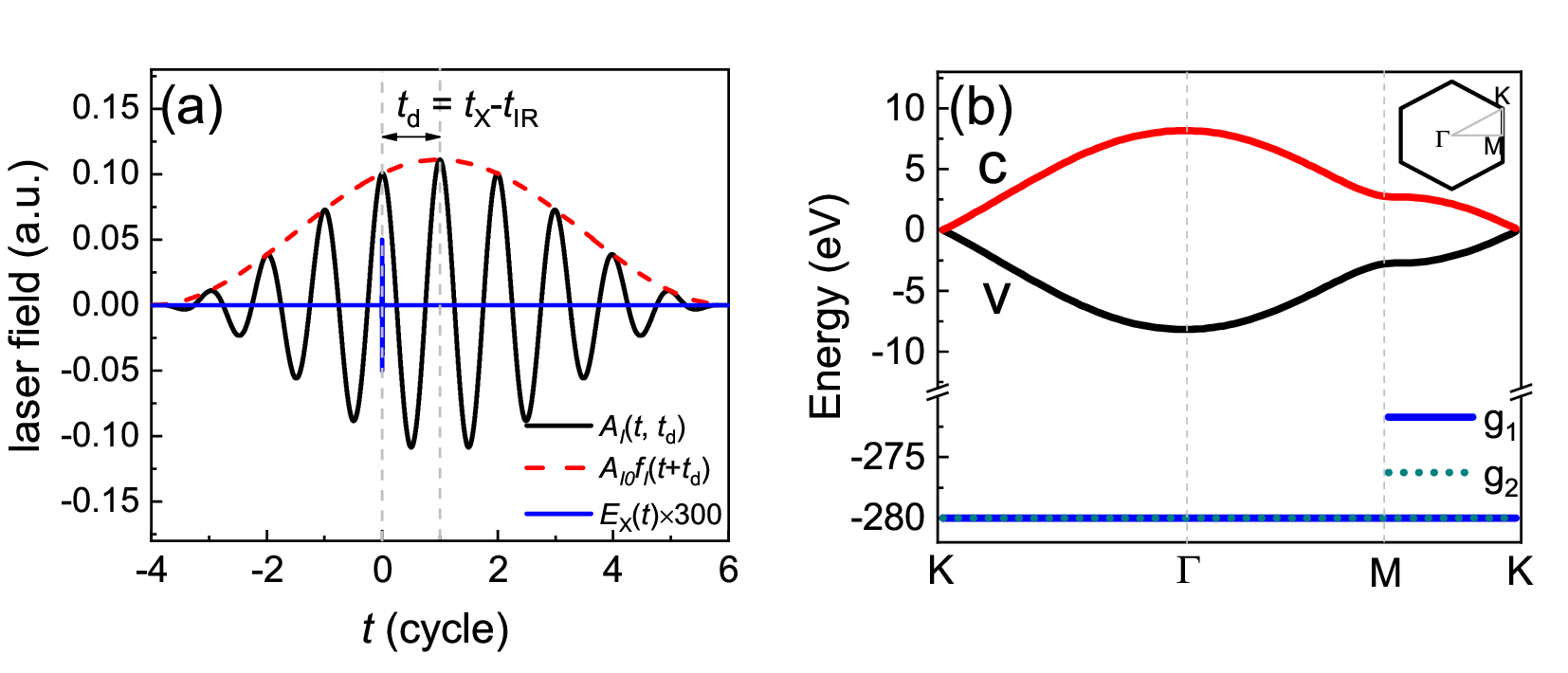}}
\caption{(a) Schematic of the time delay between the IR pump laser and X-ray probe pulse. 
(b) Two-dimensional four-band structure of graphene. 
The inset shows the first Brillouin zone of the reciprocal lattice of graphene, and three high symmetry points ($\Gamma$, $\textsc{M}$ and $\textsc{K}$) are marked.}
\label{fig:graph1}
\end{center}
\end{figure}
%%%%%%%%%%%%%%%%%%%%%%%%%%%%%%%%%%%%%%%%%%%%%%%%%%%%%

%%%%%%%%%%%%%%%%%%%%%%%%%%%%%%%%%%%%%%%%%%%%%%%%%%%%%
\begin{figure*}[t]
\begin{center}
{\includegraphics[width=18cm,height=10cm]{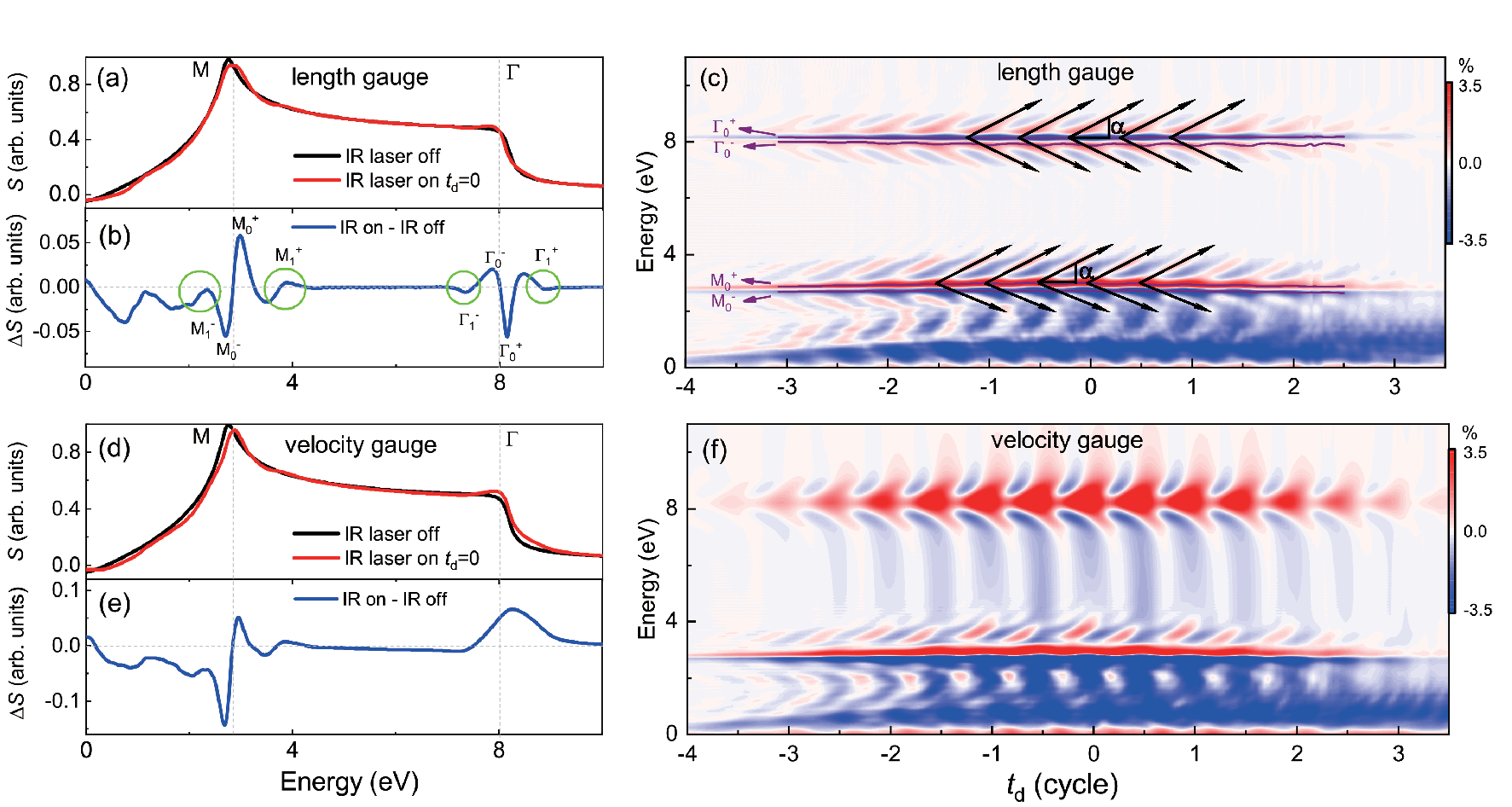}}
\caption{(a) X-ray response intensities of graphene without and with IR laser field of $t_{d} = 0$, which are calculated by Eq. (2) based on the density-matrix equations in the length gauge.
(b) Corresponding ATAS spectrum at $t_{d} = 0$ evaluated by Eq. (4).
In (b), $\textsc{M}_{0}^{-}$, $\textsc{M}_{0}^{+}$ and $\Gamma_{0}^{-}$, $\Gamma_{0}^{+}$ indicate the zeroth-order resonance peaks around $\textsc{M}$ and $\Gamma$ points, respectively, and $\textsc{M}_{1}^{-}$, $\textsc{M}_{1}^{+}$, $\Gamma_{1}^{-}$, and $\Gamma_{1}^{+}$ labeled by green rings are the first-order resonance structures.
(c) ATAS spectrum as a function of the time delay in units of IR laser optical cycles. 
In (c), the black solid arrows indicate the V-shaped structure, and $\alpha$ is the tilt angle.
The purple curves mark the zeroth-order resonance peaks that vary slowly with time delay $t_{d}$. In contrast, the first-order resonance structures periodically change with it.
Panels (d), (e) and (f) are the same as panels (a), (b) and (c), respectively, but the results are calculated by the density-matrix equations in the velocity gauge.}
\label{fig:graph1}
\end{center}
\end{figure*}
%%%%%%%%%%%%%%%%%%%%%%%%%%%%%%%%%%%%%%%%%%%%%%%%%%%%%

\subsection{Density-matrix equations in the length gauge}

We numerically simulate the ATAS spectrum of graphene in the length gauge by using the density-matrix equations in which the four energy bands have been included \cite{GioCistaro}.
Within the dipole approximation, these equations read
\begin{align}
&i \dfrac{\partial}{\partial t}\rho_{mn}(\textbf{k},t,t_{d})=[\varepsilon_{m}(\textbf{k})-\varepsilon_{n}(\textbf{k})- i \Gamma_{mn}]\rho_{mn}(\textbf{k},t,t_{d}) \nonumber\\
&+ i \textit{\textbf{E}}_{I}(t,t_{d}) \dfrac{\partial}{\partial \textbf{k}} \rho_{mn}(\textbf{k},t,t_{d}) + [\textit{\textbf{E}}_{I}(t, t_{d}) + \textit{\textbf{E}}_{X}(t)] \cdot [\hat{\textit{\textbf{D}}},\hat{\rho}]_{mn},
\end{align}
where $\Gamma_{mn}$ are the relaxation parameters.

$\textit{\textbf{E}}_{X}(t) =\textit{E}_{X} f_{X}(t) \cos(\omega_{X}t) \textit{\textbf{e}}_{z}$ is the electric field of the X-ray pulse, in which $f_{X}(t) = e^{-(4 ln 2)(t/ \tau_{X})^{2}}$ has a full width at half maximum of $\tau_{X} = 80$ attoseconds and the amplitude $\textit{E}_{X}$ corresponds to the intensity of $1 \times 10^{9}$ W/cm$^{2}$. The frequency of the X-ray pulse is $\omega_{X} = 280$ eV, which is equal to the energy gap between the Fermi surface and the core bands. 
$\textit{\textbf{e}}_{z}$ is the polarization direction, which is perpendicular to the graphene monolayer.

$\textit{\textbf{A}}_{I}(t, t_d) = \textit{A}_{I0} f_{I}(t+t_d) \cos(\omega_{I} t+\omega_{I} t_d) \textit{\textbf{e}}$ is the vector potential of the IR laser field. 
$f_{I}(t) = \cos^2 (\omega_{I}t / 2n)$ is an envelope with $n = 10$ and the amplitude of $\textit{A}_{I0}$ corresponds to a laser intensity of $1 \times 10^{11}$ W/cm$^{2}$.
$\omega_I$ is the frequency of the IR laser field, corresponding to the wavelength of $\lambda = 3000$ nm. 
$\textit{\textbf{e}}$ is the unit vector along the $\Gamma - \textsc{M}$ direction of graphene. 
The electric field of the IR laser is calculated by $\textit{\textbf{E}}_{I}(t, t_d) = - \partial \textit{\textbf{A}}_{I} (t, t_d)/ \partial t $.
As shown in Fig. 1(a), $t_d = t_{X} - t_{IR}$ is time delay, where $t_{X} = 0$ and $t_{IR}$ are the centers of the X-ray pulse and IR laser field, respectively.

For the two-dimensional four-band structure of graphene in Fig. 1(b), the interband dipole elements include $\textit{\textbf{D}}_{g_{1}g_{2}} (\textbf{k}) = \textit{\textbf{D}}_{g_{1}c} (\textbf{k}) = \textit{\textbf{D}}_{g_{2}v} (\textbf{k}) = 0$, $\textit{\textbf{D}}_{g_{1}v} (\textbf{k}) = \textit{\textbf{D}}_{g_{2}c} (\textbf{k}) = r_{z} \textit{\textbf{e}}_{z}$ and $\textit{\textbf{D}}_{cv}(\textbf{k}) = i \langle u_{c,\textbf{k}}(\textbf{r})  \vert  \triangledown_{\textbf{k}} \vert u_{v,\textbf{k}}(\textbf{r}) \rangle$.
Here, $r_{z} = \int d z \psi_{1s}^{*}(z) z \psi_{2p_{z}}(z) = 0.041$ \AA \cite{GioCistaro}, in which $\psi_{1s}(z)$ and $\psi_{2p_{z}}(z)$ are the wavefunctions of the $1s$ and $2p_{z}$ orbitals along the $z$ direction, and
$u_{c,\textbf{k}}(\textbf{r})$ [$u_{v,\textbf{k}}(\textbf{r})$] is the periodic part of the Bloch wavefunction for the conduction (valence) band of graphene \cite{GVampa,SCJiang}.

The computational complexity introduced by the gradients in Eq. (1) can be removed by transforming the crystal momentum $\textbf{k}$ into a frame moving one $\textbf{k}_t = \textbf{k} + \textit{\textbf{A}}_{I}(t, t_d)$ \cite{WVHouston}.
Under this transformation, the partial differential equation (1) reduce to ordinary differential equation,
\begin{align}
 &i \dfrac{d}{d t}\rho_{mn}(\textbf{k}_t,t,t_{d}) = [ \varepsilon_{m}(\textbf{k}_t)-\varepsilon_{n}(\textbf{k}_t)- i \Gamma_{mn}]\rho_{mn}(\textbf{k}_t,t,t_{d}) \nonumber\\
&+ [\textit{\textbf{E}}_{I}(t, t_{d}) + \textit{\textbf{E}}_{X}(t)] \cdot  \nonumber\\
&\sum_{l} [ \textit{\textbf{D}}_{ml}(\textbf{k}_t) \rho_{ln}(\textbf{k}_t,t,t_{d}) - \rho_{ml}(\textbf{k}_t,t,t_{d}) \textit{\textbf{D}}_{ln}(\textbf{k}_t)],
\end{align}
which can be readily numerically solved by the standard fourth-order Runge-Kutta algorithm.

At $t=- \infty$, electrons populate the two core bands and the valence band; thus, $\rho_{g_{1}g_{1}} (\textbf{k}_t,t=- \infty,t_{d}) =\rho_{g_{2}g_{2}}(\textbf{k}_t,t=- \infty,t_{d})=\rho_{vv}(\textbf{k}_t,t=- \infty,t_{d})=1$, and the other terms of the density matrix elements are zero.
The core-hole lifetime is set to 6.1 fs \cite{GioCistaro}; correspondingly, the relaxation parameters $\Gamma_{g_1v}=\Gamma_{g_1c}=\Gamma_{g_2v}=\Gamma_{g_2c}=0.004$ a.u., and other relaxation parameters are set to zero.

The X-ray response intensity for time delay $t_{d}$ is calculated by \cite{MetteBGaarde}
\begin{align}
S(\omega,t_{d}) = 2 \operatorname{Im}[\tilde{\mu}(\omega,t_{d}) \tilde{\textit{E}}_{X}^{*}(\omega)],  
\end{align}
where $\tilde{E}_{X}(\omega)$ is the Fourier transform of $E_{X}(t)$, and $\tilde{\textit{E}}_{X}^{*}(\omega)$ represents the complex conjugate of $\tilde{E}_{X}(\omega)$.
$\tilde{\mu}(\omega,t_{d})$ is the Fourier transform of $\mu(t,t_{d})$, which is calculated by
\begin{align}
\mu(t,t_{d}) = \sum_{\textbf{k}} \sum_{i,g}[r_{z} \rho_{ig}(\textbf{k}_t,t,t_{d}) + c.c.],
\end{align}
where $g$ represents the $g_{1}$ or $g_{2}$ band and $i$ represents the $v$ or $c$ band.

The ATAS spectrum can then be calculated according to
\begin{align}
\Delta S(\omega,t_{d}) = S(\omega,t_{d}) - S^{X}(\omega),
\end{align}
where $S^{X}(\omega)$ is the X-ray response intensity without IR laser field.

\subsection{Density-matrix equations in the velocity gauge}

One can also calculate the ATAS spectrum of graphene in the velocity gauge.
Within the dipole approximation, the corresponding density-matrix equations can be obtained \cite{Dong2}
\begin{align}
&i \dfrac{d}{d t}\rho_{mn}(\textbf{k},t,t_{d})=[\varepsilon_{m}(\textbf{k})-\varepsilon_{n}(\textbf{k})- i \Gamma_{mn}]\rho_{mn}(\textbf{k},t,t_{d}) \nonumber\\
& + [\textit{\textbf{A}}_{I}(t, t_{d}) + \textit{\textbf{A}}_{X}(t)] \cdot [\hat{\textbf{p}}(\textbf{k}),\hat{\rho}]_{mn},
\end{align}
where $\textit{\textbf{A}}_{X}(t)  = - \int^{t}_{- \infty} \textit{\textbf{E}}_{X}(t^{\prime}) d t^{\prime}$ is the vector potential of the X-ray pulse.
The momentum dipole elements consist of the intraband dipole elements $\textbf{p}_{cc}(\textbf{k})= \triangledown_{\textbf{k}} \varepsilon_{c}(\textbf{k}) =- \textbf{p}_{vv}(\textbf{k})$ and $\textbf{p}_{g_{1}g_{1}}(\textbf{k})=\textbf{p}_{g_{2}g_{2}}(\textbf{k}) = 0$ and the interband dipole elements
$\textbf{p}_{m\bar{m}}(\textbf{k}) =i (\varepsilon_{m}(\textbf{k})-\varepsilon_{\bar{m}}(\textbf{k})) \textbf{D}_{m\bar{m}}(\textbf{k})$, where $\bar{m} \ne m$.

\subsection{ATAS spectrum from numerically solving density-matrix equations}

%%%%%%%%%%%%%%%%%%%%%%%%%%%%%%%%%%%%%%%%%%%%%%%%%%%%%%%
\begin{figure}[t]
\begin{center}
{\includegraphics[width=8.5cm,height=9.5cm]{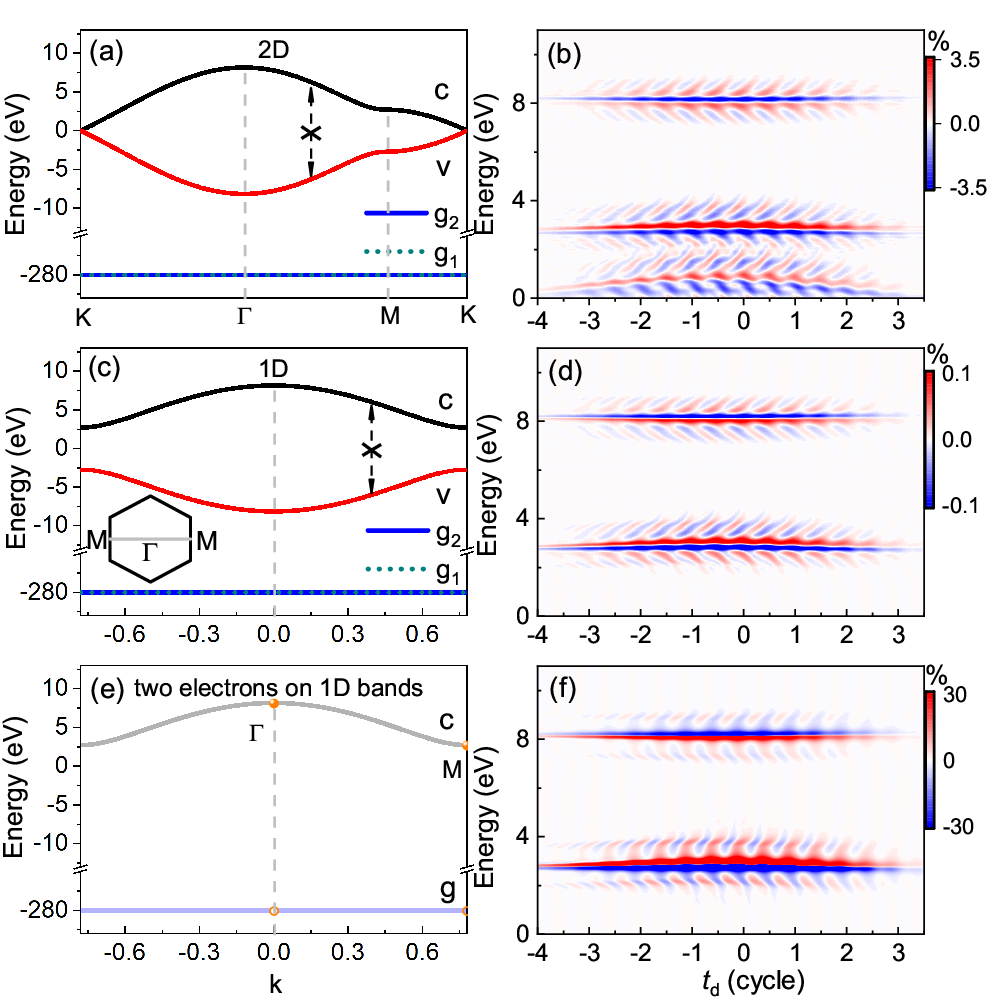}}
\caption{(a) Schematic of the two-dimensional (2D) four-band model in which the transition between the valence and conduction bands has been artificially blocked. 
(b) ATAS spectrum as a function of the time delay in units of IR laser optical cycles, which is calculated using density-matrix equation (2) based on the model in (a). 
(c) One-dimensional (1D) four-band model obtained by taking a section along the $\Gamma - \textsc{M}$ direction of the 2D energy bands in (a), as presented by the gray line of the inset, where the transition between valence and conduction bands has also been blocked.
(d) ATAS spectrum corresponding to the model in (c).  
(e) Simplified model which includes two electrons.
The lattice momenta of the electrons are $\texttt{k} = 0$ ($\Gamma$ point) and $\texttt{k} \approx 0.78$ a.u. ($\textsc{M}$ point).
(f) ATAS spectrum based on the simplified model in (e).}
\label{fig:graph1}
\end{center}
\end{figure}
%%%%%%%%%%%%%%%%%%%%%%%%%%%%%%%%%%%%%%%%%%%%%%%%%%%%%

By numerically solving the above four-band density-matrix equations, we calculate the ATAS spectrum in both the length and velocity gauges.

Figure 2(a) shows the X-ray response intensities of graphene without and with an IR laser field of $t_{d} = 0$, which are calculated by Eq. (3) based on the density-matrix equations in the length gauge.
The corresponding ATAS spectrum at $t_{d} = 0$ evaluated by Eq. (5) is presented in Fig. 2(b).
Figure 2(c) shows the ATAS spectrum as a function of the time delay in units of IR laser optical cycles.
In the absorption spectrum in Fig. 2(b), we observe interesting zeroth-order resonant peaks $\textsc{M}_{0}^{-}$ and $\textsc{M}_{0}^{+}$ (or $\Gamma_{0}^{-}$ and $\Gamma_{0}^{+}$) around $\textsc{M}$ (or $\Gamma$) point, and they vary slowly with time delay $t_{d}$ as marked by the purple curves in Fig. 2(c).
In contrast, the first-order resonant structures $\textsc{M}_{1}^{-}$ and $\textsc{M}_{1}^{+}$, $\Gamma_{1}^{-}$ and $\Gamma_{1}^{+}$ labeled by green rings in Fig. 2(b) periodically vary with time delay $t_{d}$, as shown in Fig. 2(c).
In Fig. 2(c), apparent “fish bone” structures consisting of the zeroth- and first-order resonant structures around the $\textsc{M}$ point (near $2.72$ eV) and $\Gamma$ point (approximately $8.16$ eV) can be observed.
The black solid arrows indicate the V-shaped structure (corresponding to local maximums) with a tilt angle of $\alpha$ in the “fish bone” structure, which has a period that is half the IR laser optical period.

Our calculated ATAS spectrum shown in Fig. 2(c) is analogous to that of Ref. \cite{GioCistaro}, in which the pulse duration of the IR laser is three optical cycles. 
The “fish bone” structure has also been observed in an ab initio simulation of the ATAS spectrum of monolayer hexagonal boron nitride in \cite{ShunsukeASato}. 
The periodicity of the V-shaped structure was found to emerge in other materials, such as diamond and GaAs \cite{MLucchini,FSchlaepfer}, which was attributed to the dynamical Franz-Keldysh effect related to the intraband motion of electrons.
Despite these experiments, simulations and theoretical investigations, an explicit mechanism of the “fish bone” resonance structure in an analytical form is still lacking.

Figures 2(d), 2(e) and 2(f) show the results calculated by using the four-band density-matrix equations in the velocity gauge.
By comparing Figs. 2(b) and 2(e), we can see that the ATAS spectra near the $\textsc{M}$ point are qualitatively consistent; however, a large discrepancy exists near the $\Gamma$ point, leading to the different “fish bone” structures near the $\Gamma$ point in Figs. 2(c) and 2(f).

In principle, the ATAS results should be gauge-independent.
The difference is due to the four-band approximation.
To obtain more accurate ATAS spectrum near the $\Gamma$ point, one should consider more conduction bands in the density-matrix equations in the velocity gauge \cite{WELamb,MScully}.
In contrast, the density-matrix equations in the length gauge describe the electron dynamics using time-dependent Houston states, which are best thought of as an adiabatic basis \cite{MWu}.
Our IR laser with a wavelength of $3000$ nm and a pulse duration of ten cycles satisfies the requirements of the adiabatic approximation.
In the following, we investigate the underlying mechanism of the ATAS spectrum mainly based on the length gauge.

\section{Analytical investigation of “fish bone” resonance structure in the ATAS spectrum}
\label{s3}

\subsection{Simplified model}
We first study the influence of intraband and interband transitions on the ATAS spectrum.
In the two-dimensional four-band model in Fig. 1(b), we block the interband transition between the valence and conduction bands, as shown in Fig. 3(a).
The corresponding ATAS spectrum is presented in Fig. 3(b).
By comparing Fig. 3(b) and Fig. 2(c), one can obtain that the interband transition has a significant effect on the ATAS spectrum near the $\textsc{K}$ point; however, it plays a small role in the spectrum at the $\Gamma$ and $\textsc{M}$ points.
The underlying mechanism is that for the band model in Fig. 1(b), the electrons near the $\textsc{K}$ point can be easily excited from the valence band to the conduction band by the IR laser.
This process blocks the transition of electrons from the $g_{2}$ band to the $c$ band under the X-ray pulse excitation and therefore results in an absorption decrease near the $\textsc{K}$ point, as shown in Fig. 2.
At the $\textsc{M}$ and $\Gamma$ points, however, the wider energy gaps block interband transition process caused by the IR laser, and the generation mechanism of the ATAS spectrum arises from the intraband dynamics of electrons.
In the following work, we mainly investigate the mechanism of the “fish bone” structure near $\textsc{M}$ and $\Gamma$ points.

We further simplify our two-dimensional model in Fig. 3(a) to the one-dimensional model in Fig. 3(c) by taking a section along the $\Gamma - \textsc{M}$ direction of the 2D band structure.
The dispersion relations of the one-dimensional bands are $\epsilon_{c}(\texttt{k}) = - \epsilon_{v}(\texttt{k}) = \varepsilon_{c}(\texttt{k} = \texttt{k}_{x}, \texttt{k}_{y} = 0) =  \gamma_{0} \sqrt{5+ 4\cos(3 \texttt{k} d/2)}$ and $\epsilon_{g} = \epsilon_{g1} (\texttt{k}) =\epsilon_{g2}(\texttt{k}) =-280$ eV.
Based on this model, the corresponding ATAS spectrum is presented in Fig. 3(d).
One can find that the ATAS near the $\textsc{K}$ point disappears, while the “fish bone” structures around the $\textsc{M}$ and $\Gamma$ points are well consistent with those in Fig. 3(b). 
Therefore, we can exploit the one-dimensional model in Fig. 3(c) to study the “fish bone” structure.

Furthermore, in the band model in Fig. 3(c), the electrons in the $g_{1}$ and $v$ bands cannot jump to the $g_{2}$ or $c$ bands because $\textit{\textbf{D}}_{g_{1}c} (\texttt{k}) = \textit{\textbf{D}}_{vc}(\texttt{k}) = \textit{\textbf{D}}_{g_{1}g_{2}} (\texttt{k}) = \textit{\textbf{D}}_{vg_{2}}(\texttt{k}) = 0$.
Additionally, although $\textit{\textbf{D}}_{g_{1}v} (\texttt{k}) \neq 0 $, the transition between the $g_{1}$ and $v$ bands is forbidden because $\rho_{g_{1}g_{1}} (\texttt{k}_t, t, t_{d}) = \rho_{vv}(\texttt{k}_t, t, t_{d})=1$.
Therefore, the one-dimensional four-band model in Fig. 3(c) is equivalent to the one-dimensional two-band structure in Fig. 3(e) consisting of $g \equiv g_{2}$ and $c$ bands. 
We propose the simplified model shown in Fig. 3(e), which includes two electrons with lattice momenta of $\texttt{k} = 0$ ($\Gamma$ point) and $\texttt{k} \approx 0.78$ a.u. ($\textsc{M}$ point), and the corresponding ATAS spectrum is shown in Fig. 3(f).
The “fish bone” structures of ATAS spectrum in Fig. 3(f) are qualitatively consistent with those of Fig. 3(d).
In the following, based on this simplified model in Fig. 3(e), we develop an analytical theory to investigate the underlying mechanism of the “fish bone” structure.

\subsection{Analytic deduction of ATAS spectrum $\Delta S(\omega,t_{d})$}
Based on the simplified model in Fig. 3(e), we deduce an analytical formulation of the ATAS spectrum.
Because the X-ray pulse is relatively short and weak, it can be approximated to a $\delta$ function $E_{X}(t) =E_{X} \delta(t)$.
The electrons can be instantaneously excited from the $g$ band to the $c$ band by the X-ray pulse at the moment of $t = 0$.
According to perturbation theory and Eq. (2), the density matrix elements change from $\rho_{gg} (\texttt{k}_t,t < 0^{-},t_{d}) = 1$, $\rho_{cc} (\texttt{k}_t,t < 0^{-},t_{d}) = 0$, and $\rho_{cg} (\texttt{k}_t,t < 0^{-},t_{d}) = 0$ to $\rho_{gg} (\texttt{k}_t,t = 0^{+},t_{d}) \approx 1$, $\rho_{cc} (\texttt{k}_t,t = 0^{+},t_{d}) \approx 0$, and $\rho_{cg} (\texttt{k}_t,t = 0^{+},t_{d}) \approx -i \textit{E}_{X} r_{z}$.
Next, the time-dependent evolution of density matrix elements is dominated only by the IR laser, and one can obtain $\rho_{cg} (\texttt{k}_t,t > 0^{+}, t_{d}) = -i \textit{E}_{X} r_{z} e^{-i \int^{t}_{0} ( \epsilon_{c}(\texttt{k}+A_{I}(t^{\prime}, t_{d}) ) - \epsilon_{g})d t^{\prime}} e^{- \Gamma_{0} t}$. 
According to Eq. (4), when $t < 0^{-}$, the time-dependent dipole is $\mu_{\texttt{k}} (t, t_{d}) = 0$, and when $t > 0^{+}$, it is $\mu_{\texttt{k}} (t, t_{d}) = - 2 \textit{E}_{X} r_{z}^{2} \sin[\int^{t}_{0}[\epsilon_{c}(\texttt{k}+A_{I}(t^{\prime},t_{d})) - \epsilon_{g} ]d t^{\prime}] e^{- \Gamma_{0} t}$. Here relaxation parameter $\Gamma_{0}=0.004$ a.u. is consistent with that used in the numerical calculation.

According to Eqs. (3) and (4), the response intensity is calculated by 
$S(\omega, t_{d}) = \sum_{\texttt{k}} S_{\texttt{k}}(\omega, t_{d})$ and $S_{\texttt{k}}(\omega, t_{d}) = 2 \operatorname{Im}[\tilde{\mu}_{\texttt{k}}(\omega, t_{d}) \tilde{E}_{X}^{*}(\omega)] 
\propto  \operatorname{Im}[\tilde{\mu}_{\texttt{k}}(\omega, t_{d})] = \operatorname{Im}[\int_{0}^{\infty} \mu^{X}_{\texttt{k}}(t) e^{-i \omega t} d t]$. 
When the IR laser is off, the response intensity is $S^{X}_{\texttt{k}}(\omega) \propto  \dfrac{\Gamma_{0}}{\Gamma_{0}^{2}+[\omega-\epsilon_c(\texttt{k})]^{2}} = L[\omega,\epsilon_c(\texttt{k})]$, in which $\texttt{k} = \texttt{k}_{\Gamma}=0$ or $\texttt{k} = \texttt{k}_{\textsc{M}} \approx 0.78$.
$L(\omega,x) = \dfrac{\Gamma_{0}}{\Gamma_{0}^{2}+(\omega-x)^{2}}$ is the Lorentzian line shape centered at $x$. (See Appendix A for the detailed derivation. Note that the response intensity spectra have been shifted by $\epsilon_{g}$ in the energy domain.)

When the IR laser is turned on, the response intensity of the electron at the $\Gamma$ or $\textsc{M}$ point can be evaluated by $S_{\texttt{k}}(\omega,t_{d})$.
The analytical expression of the ATAS in general takes following form:
\begin{align}
\Delta S_{\texttt{k}} (\omega,t_{d}) & = S_{\texttt{k}}(\omega,t_{d}) - S^{X}_{\texttt{k}}(\omega) \nonumber\\
& = \Delta S_{\texttt{k}}^{0} (\omega,t_{d}) + \Delta S_{\texttt{k}}^{1} (\omega,t_{d}),
\end{align}
where $\Delta S_{\texttt{k}}^{0} (\omega,t_{d})$ and $\Delta S_{\texttt{k}}^{1} (\omega,t_{d})$ are the zeroth- and first-order resonance structures, respectively.
We can obtain 
\begin{align}
\Delta S_{\texttt{k}}^{0} (\omega,t_{d}) = J_{0}[b_{\texttt{k}}(t_{d})] \cdot L[\omega,\epsilon_{s}(\texttt{k},t_{d})] - L[\omega,\epsilon_c(\texttt{k})]
\end{align}
and
\begin{align}
\Delta S_{\texttt{k}}^{1} (\omega , t_{d}) & = J_{1}[b_{\texttt{k}}(t_{d})] \cdot L\left[\omega,\epsilon_{s}(\texttt{k},t_{d})+2 \omega_{I}\right] \cos (2 \omega_{I} t_{d}) \nonumber\\
&+ J_{1}[b_{\texttt{k}}(t_{d})] \cdot F\left[\omega,\epsilon_{s}(\texttt{k},t_{d})+2 \omega_{I}\right] \sin (2 \omega_{I} t_{d}) \nonumber\\
&- J_{1}[b_{\texttt{k}}(t_{d})] \cdot L\left[\omega,\epsilon_{s}(\texttt{k},t_{d})-2 \omega_{I}\right] \cos (2 \omega_{I} t_{d}) \nonumber\\
&+ J_{1}[b_{\texttt{k}}(t_{d})] \cdot F\left[\omega,\epsilon_{s}(\texttt{k},t_{d})-2 \omega_{I}\right] \sin (2 \omega_{I} t_{d}),
\end{align}
where $F(\omega,x) = \dfrac{\omega - x}{\Gamma_{0}^{2} + (\omega - x)^{2}}$ is the Fano line shape \cite{UFANO,ChristianOtt} centered at $x$.
Here, we define $\epsilon_{s}(\texttt{k},t_{d}) = \epsilon_{c}(\texttt{k})  + A_{I0}^{2} f_{I}^{2}(t_{d}) / (4 m_{\texttt{k}}^{*})$ and $b_{\texttt{k}}(t_{d}) = \textit{A}_{I0}^{2}  f_{I}^{2}(t_{d}) / (8 \omega_{I} m_{\texttt{k}}^{*})$, where $ m_{\texttt{k}}^{*}= 1 / \nabla_{\texttt{k}}^{2} \epsilon_{c}(\texttt{k})$ is the effective mass.
$J_{n}(x)$ is the $n$th-order Bessel function.

%%%%%%%%%%%%%%%%%%%%%%%%%%%%%%%%%%%%%%%%%%%%%%%%%%%%%%
\begin{figure}[t]
\begin{center}
{\includegraphics[width=8.5cm,height=12cm]{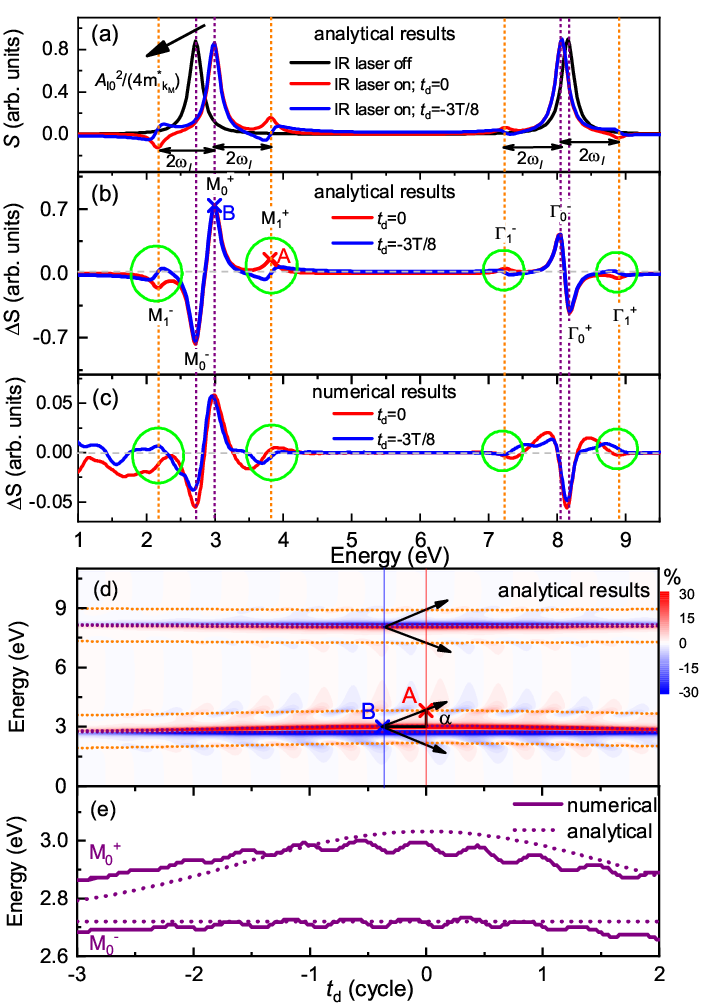}}
\caption{(a) The analytical response intensities $S^{X}_{\texttt{k}}(\omega)$, $S_{\texttt{k}}(\omega,t_{d}=0)$ and $S_{\texttt{k}}(\omega,t_{d}=-3T/8)$.
(b) Corresponding absorption spectra $\Delta S_{\texttt{k}}(\omega,t_{d}=0)$ and $\Delta S_{\texttt{k}}(\omega,t_{d}=-3T/8)$.
(c) The numerical ATAS spectra for time delays $t_{d}=0$ and $t_{d}=-3T/8$, which are extracted from Fig. 2(c).
In (a), (b), and (c), the vertical purple dotted lines mark the zeroth-order resonance peaks $\textsc{M}_{0}^{-}$, $\textsc{M}_{0}^{+}$, $\Gamma_{0}^{-}$ and $\Gamma_{0}^{+}$ that correspond to $\Delta S_{\texttt{k}}^{0}(\omega,t_{d})$, and the orange dotted lines are the centers of first-order resonance structures $\textsc{M}_{1}^{-}$, $\textsc{M}_{1}^{+}$, $\Gamma_{1}^{-}$ and $\Gamma_{1}^{+}$ labeled by green rings, corresponding to $\Delta S_{\texttt{k}}^{1}(\omega,t_{d})$.
(d) The analytical ATAS spectra $\Delta S_{\texttt{k}}(\omega,t_{d})$ calculated by Eq. (7).
In (d), the black arrows indicate the V-shaped structure, whose tilt angle is defined as $\alpha$.
Around the $\textsc{M}$ and $\Gamma$ points, the horizontal purple dotted lines are $\epsilon_{c}(\texttt{k})$ and $\epsilon_{s}(\texttt{k},t_{d})$, and orange dotted lines correspond to $\epsilon_{s}(\texttt{k},t_{d}) \pm 2\omega_{I}$.
In (b) and (d), “A” and “B” points labeled by red and black crosses are local maximum values.
In (e), the purple solid curves reproduce $\textsc{M}_{0}^{-}$ and $\textsc{M}_{0}^{+}$ in Fig. 2(c), and the dotted lines are the corresponding analytical results $\epsilon_{c}(\texttt{k}_{\textsc{M}})$ and $\epsilon_{s}(\texttt{k}_{\textsc{M}},t_{d})$.
}
\label{fig:graph1}
\end{center}
\end{figure}
%%%%%%%%%%%%%%%%%%%%%%%%%%%%%%%%%%%%%%%%%%%%%%%%%%%%%%%%

\subsection{Resonance peaks in the ATAS spectrum}
When the IR laser is off, the response intensity $S_{\texttt{k}}^{X}(\omega) = L[\omega,\epsilon_{c}(\texttt{k})]$ (black line) is shown in Fig. 4(a), which presents two Lorentzian line shapes whose peaks are located at $\epsilon_{c}(\texttt{k}_{\textsc{M}}) \approx 2.72$ eV and $\epsilon_{c}(\texttt{k}_{\Gamma}) \approx 8.16$ eV.

When the IR laser of $t_{d} = 0$ is turned on, the response intensity (red line in Fig. 4(a)) is $S_{\texttt{k}}(\omega, t_{d} = 0) = J_{0}(b_{\texttt{0k}}) L[\omega, \epsilon_{\texttt{0}s}(\texttt{k})] +J_{1}(b_{\texttt{0k}}) L\left[\omega, \epsilon_{\texttt{0}s}(\texttt{k})+2 \omega_{I}\right] - J_{1}(b_{\texttt{0k}}) L\left[\omega, \epsilon_{\texttt{0}s}(\texttt{k})-2 \omega_{I}\right]$, where $\epsilon_{\texttt{0}s}(\texttt{k}) =\epsilon_{s}(\texttt{k},t_{d}=0) = \epsilon_{c}(\texttt{k})  + A_{I0}^{2} / (4 m_{\texttt{k}}^{*})$ and
$b_{\texttt{0k}} = b_{\texttt{k}}(t_{d}=0) = \textit{A}_{I0}^{2} / (8 \omega_{I} m_{\texttt{k}}^{*})$. 
Comparing with $S_{\texttt{k}}^{X}(\omega)$, the zeroth-order resonance peaks associated with $J_{0}(b_{\texttt{0k}})$ term are shifted by $A_{I0}^{2}/(4  m_{\texttt{k}}^{*})$.
For the electron of the $\textsc{M}$ point (or the $\Gamma$ point) whose effective mass $m_{\texttt{k}_{\textsc{M}}}^{*}$ (or $m_{\texttt{k}_{\Gamma}}^{*}$) is $0.3$ a.u. (or $-0.9$ a.u.), the energy shift $A_{I0}^{2} / (4 m_{\texttt{k}_{\textsc{M}}}^{*})$ (or $A_{I0}^{2} / (4 m_{\texttt{k}_{\Gamma}}^{*})$) is equal to $0.27$ eV (or $-0.09$ eV).
In addition, when IR laser of $t_{d} = 0$ is on, for each electron with lattice momentum $\texttt{k}_{\textsc{M}}$ or $\texttt{k}_{\Gamma}$, two additional first-order resonance structures associated with $J_{1}(b_{\texttt{0k}})$ terms appear, which exhibit Lorentzian line shape.
The energy intervals between the zeroth-order resonance peak and two first-order resonance structures are $2 \omega_{I}$, as shown in Fig. 4(a).

In Fig. 4(a), the blue line shows the response intensity $S_{\texttt{k}}(\omega, t_{d} = -3T/8) = J_{0}(b_{\texttt{1k}}) L[\omega, \epsilon_{\texttt{1}s}(\texttt{k})] +J_{1}(b_{\texttt{1k}}) F\left[\omega, \epsilon_{\texttt{1}s}(\texttt{k})+2 \omega_{I}\right] + J_{1}(b_{\texttt{1k}}) F\left[\omega, \epsilon_{\texttt{1}s}(\texttt{k})-2 \omega_{I}\right]$, where $\epsilon_{\texttt{1}s}(\texttt{k}) =\epsilon_{s}(\texttt{k},t_{d}=-3T/8) \approx \epsilon_{\texttt{0}s}(\texttt{k})$ and $b_{\texttt{1k}} = b_{\texttt{k}}(t_{d}=-3T/8) \approx b_{\texttt{0k}}$ for both $\textsc{M}$ and $\Gamma$ points. 
In contrast to $S_{\texttt{k}}(\omega, t_{d} = 0)$, the first-order resonance structures of response intensity $S_{\texttt{k}}(\omega, t_{d} = -3T/8)$ exhibit the Fano line shape.

Figure 4(b) shows analytical ATAS spectra $\Delta S_{\texttt{k}}(\omega, t_{d} = 0)$ and $\Delta S_{\texttt{k}}(\omega, t_{d} = -3T/8)$ calculated by Eq. (7).
One can find two zeroth-order peaks $\textsc{M}_{0}^{-}$ and $\textsc{M}_{0}^{+}$ (or $\Gamma_{0}^{-}$ and $\Gamma_{0}^{+}$) corresponding to $\Delta S_{\texttt{k}}^{0}(\omega, t_{d})$, as well as two first-order resonance structures $\textsc{M}_{1}^{-}$ and $\textsc{M}_{1}^{+}$ (or $\Gamma_{1}^{-}$ and $\Gamma_{1}^{+}$) labeled by green rings corresponding to $\Delta S_{\texttt{k}}^{1}(\omega, t_{d})$ around the $\textsc{M}$ (or $\Gamma$) point.

In Fig. 4(c), we show numerical ATAS spectra for time delays $t_{d} = 0$ and $t_{d} = -3T/8$, which are extracted from Fig. 2(c).
Comparing the results in Fig. 4(b) and Fig. 4(c), one can obtain that the zeroth- and first-order resonance structures are qualitatively consistent.
Quantitatively, there are some deviations between the analytical and numerical results, especially for the first-order resonance structures, which arise from the fact that our analytical results are based on the simplified model that only considers the electrons of $\Gamma$ and $\textsc{M}$ points in the Brillouin zone.

\subsection{V-shaped structure in the ATAS spectrum}

The analytical ATAS spectra as a function of the time delay, which are calculated by Eq. (7), are shown in Fig. 4(d).
The black arrows indicate the V-shaped structure that corresponds to local maximum ATAS spectrum amplitudes.
According to $\Delta S_{\texttt{k}}^{0}(\omega, t_{d})$ of Eq. (8), one can obtain that the zeroth-order resonance peaks vary slowly with time delay $t_{d}$, as shown in Fig. 4(d).
In contrast, $\Delta S_{\texttt{k}}^{1}(\omega, t_{d})$ of Eq. (9) implies that as the time delay continuously varies, the first-order resonance structures periodically present  Lorentzian or Fano line shapes, forming the V-shaped structure in the ATAS spectrum.
Corresponding to $\cos (2 \omega_{I} t_{d})$ and $\sin (2 \omega_{I} t_{d})$, the period of the V-shaped structure is half the IR laser optical period, as shown in Figs. 4(d) and 2(c).
The zeroth and first-order resonance structures make up the “fish bone” structure in the ATAS spectrum.

We define the tilt angle $\alpha$ of the V-shaped structure in Fig. 4(d).
The horizontal purple ($\epsilon_{c}(\texttt{k})$, $\epsilon_{s}(\texttt{k},t_{d})$) and orange dotted lines ($\epsilon_{s}(\texttt{k},t_{d}) \pm 2\omega_{I}$) in Fig. 4(d) indicate the centres of the zeroth and first-order resonance structures.
The vertical red and blue lines mark the time delays $t_{d} = 0$ and $t_{d} = -3T/8$, and corresponding ATAS spectra have been shown in Fig. 4(b).
As shown in Figs. 4(d) and 4(b), the “A” points labeled by red crosses are local maximal values, which satisfy both $\frac{\partial}{\partial \omega} \Delta S(\omega,t_{d}) |_{t_{d} = 0}= 0$ and $\frac{\partial}{\partial t_{d}} \Delta S(\omega,t_{d}) |_{\omega = \epsilon_{s}(\texttt{k}_{\textsc{M}},t_{d}) + 2 \omega_{I}} = 0$.
The “B” points labeled by black crosses satisfy $\frac{\partial}{\partial \omega} \Delta S(\omega,t_{d}) |_{t_{d} = -3T/8} = 0$ and are located at the zeroth-order peak $\textsc{M}_{0}^{+}$, as shown in Fig. 4(b).
The energy and time intervals between points “A” and “B” are $2 \omega_{I}$ and $3T/8$, respectively.
Therefore, in Fig. 4(d), the tilt angle $\alpha$ of the V-shaped structure is defined by $\tan(\alpha) \approx \frac{2 \omega_{I}}{3T/8} = \frac{8 \omega_{I}^{2}}{3 \pi}$.
This implies that the tilt angle of the V-shaped structure increases with the IR laser frequency.

In Fig. 4(e), the two purple solid curves $\textsc{M}_{0}^{-}$ and $\textsc{M}_{0}^{+}$ reproduce the zeroth-order resonance peaks of the “fish bone” structure near the $\textsc{M}$ point in Fig. 2(c).
The two purple dotted lines are the analytical results $\epsilon_{c}(\texttt{k}_{\textsc{M}})$ and $\epsilon_{s}(\texttt{k}_{\textsc{M}},t_{d})$, corresponding to $\textsc{M}_{0}^{-}$ and $\textsc{M}_{0}^{+}$, respectively.
One can obtain that for different time delays, our analytical theory can qualitatively predict the energy shifts of the zeroth-order fringes of the “fish bone” structure in the ATAS spectrum.

\section{Conclusion}
\label{s4}
In summary, we investigate the ATAS spectrum of graphene by numerically solving four-band density-matrix equations in both the length and velocity gauges, which shows apparent “fish bone” resonance structures. 
To gain insight into these interesting structures, we develop a simplified model that only considers the electrons of the $\Gamma$ and $\textsc{M}$ points.
With the help of this model, we can obtain an analytical expression of the ATAS spectrum that consists of the zeroth- and first-order resonance structures associated with Bessel functions.
Our theory predicts that the zeroth-order fringe of the “fish bone” structure slowly varies with the time delay, while the first-order fringes alternately change between the Lorentzian line shape and Fano line shape at twice the IR laser frequency, leading to the “fish bone” resonance structure.
Our analytical results of the "fish bone" resonance structure in the ATAS spectrum are rather general and can be applicable to other two-dimensional materials or even bulk solids, which might stimulate experiments in the field.

\section*{ACKNOWLEDGMENTS}

This work is supported by NSAF (Grant No. U1930403). We acknowledge valuable discussions with Professor Difa Ye.

\appendix

\section{Analytical deduction of the ATAS spectrum based on the simplified model}

We deduce the analytical results of the ATAS spectrum based on the simplified model shown in Fig. 3(e).
Because the X-ray pulse is relatively short and weak, it can be approximated to a $\delta$ function $E_{X}(t) =E_{X} \delta(t)$.
In the simplified model, the electrons can be instantaneously excited from the $g$ band to the $c$ band by the X-ray pulse.
According to perturbation theory and Eq. (2), the density matrix elements change from $\rho_{gg} (\texttt{k}_t,t < 0^{-},t_{d}) = 1$, $\rho_{cc} (\texttt{k}_t,t < 0^{-},t_{d}) = 0$, and $\rho_{cg} (\texttt{k}_t,t < 0^{-},t_{d}) = 0$ to $\rho_{gg} (\texttt{k}_t,t = 0^{+},t_{d}) \approx 1$, $\rho_{cc} (\texttt{k}_t,t = 0^{+},t_{d}) \approx 0$, and $\rho_{cg} (\texttt{k}_t,t = 0^{+},t_{d}) \approx -i \textit{E}_{X} r_{z}$.
Next, the time-dependent evolution of density matrix elements is dominated only by IR laser, and one can obtain $\rho_{cg} (\texttt{k}_t,t > 0^{+}, t_{d}) = -i \textit{E}_{X} r_{z} e^{-i \int^{t}_{0} ( \epsilon_{c}(\texttt{k}+A_{I}(t^{\prime}, t_{d}) ) - \epsilon_{g})d t^{\prime}} e^{- \Gamma_{0} t} $. 
According to Eq. (4), when $t < 0^{-}$, the time-dependent dipole is $\mu_{\texttt{k}} (t, t_{d}) = 0$, and when $t > 0^{+}$, it is 
\begin{align}
&\mu_{\texttt{k}} (t, t_{d}) = - 2 \textit{E}_{X} r_{z}^{2} \sin[\int^{t}_{0}[\epsilon_{c}(\texttt{k}+A_{I}(t^{\prime},t_{d})) - \epsilon_{g} ]d t^{\prime}] e^{- \Gamma_{0} t}.
\end{align}

According to Eqs. (3) and (4), the response intensity is calculated by 
$S(\omega, t_{d}) = \sum_{\texttt{k}} S_{\texttt{k}}(\omega, t_{d})$ and $S_{\texttt{k}}(\omega, t_{d}) = 2 \operatorname{Im}[\tilde{\mu}_{\texttt{k}}(\omega, t_{d}) \tilde{E}_{X}^{*}(\omega)]  
\propto  \operatorname{Im}[\tilde{\mu}_{\texttt{k}}(\omega, t_{d})]$. 
When the IR laser is off, the time-dependent dipole is $\mu^{X}_{\texttt{k}} (t) = - 2 E_{X} r_{z}^{2} \sin[(\epsilon_{c}(\texttt{k}) - \epsilon_{g}) t] e^{- \Gamma_{0} t}$ for $t > 0^{+}$. 
The response intensity is
\begin{align}
S^{X}_{\texttt{k}}(\omega) & \propto  \operatorname{Im}[\tilde{\mu}^{X}_{\texttt{k}}(\omega)] = \operatorname{Im}[\int_{0}^{\infty} \mu^{X}_{\texttt{k}}(t) e^{-i \omega t} d t]  \nonumber\\
& \propto \dfrac{\Gamma_{0}}{\Gamma_{0}^{2}+(\omega-[\epsilon_c(\texttt{k})-\epsilon_{g}])^{2}}. 
\end{align}

When the IR laser is turned on, the time-dependent dipole of Eq. (A1) can be deduced to be
\begin{align}
\mu_{\texttt{k}}&(t, t_{d}) = - 2 \textit{E}_{X} r_{z}^{2} \sin[\epsilon_{c}(\texttt{k})t - \epsilon_{g} t  \nonumber\\
&+ \nabla_{\texttt{k}} \epsilon_{c}(\texttt{k}) \int_{0}^{t} A_{I}(t^{\prime},t_{d})dt^{\prime} + \frac{1}{2} \nabla^{2}_{\texttt{k}} \epsilon_{c}(\texttt{k}) \int_{0}^{t} A_{I}^{2}(t^{\prime},t_{d})dt^{\prime}  \nonumber\\
&+ \frac{1}{6}  \nabla^{3}_{\texttt{k}} \epsilon_{c}(\texttt{k}) \int_{0}^{t} A_{I}^{3}(t^{\prime},t_{d})dt^{\prime} + \cdots] e^{- \Gamma_{0} t}.
\end{align}

For both the $\Gamma$ and $\textsc{M}$ points in the one-dimensional two-band structure, one can obtain $\nabla_{\texttt{k}} \epsilon_{c}(\texttt{k}) = 0$ and $\nabla^{3}_{\texttt{k}} \epsilon_{c}(\texttt{k}) = 0$, and we ignore the higher-order terms of Eq. (A3). 
To simplify the integral with respect to time in Eq. (A3), we consider the vector potential $A_{I}(t, t_{d}) = A_{I0} f_{I} (t + t_{d}) \cos(\omega_{I}t + \omega_{I} t_{d}) \approx \textit{A}_{I0} f_{I}(t_{d}) \cos(\omega_{I}t + \varphi) $, with $\varphi = \omega_{I} t_{d}$, and the time-dependent dipole can be approximated as
\begin{align}
\mu_{\texttt{k}} (t, t_{d}) \approx &- 2 \textit{E}_{X} r_{z}^{2} \sin[\epsilon_{c}(\texttt{k})t - \epsilon_{g} t  + \frac{1}{2}  \nabla^{2}_{\texttt{k}} \epsilon_{c}(\texttt{k})  \textit{A}_{I0}^{2} \nonumber\\
& \cdot f_{I}^{2}(t_{d}) \int_{0}^{t} \dfrac{\cos(2 \omega_{I} t^{\prime} + 2 \varphi) + 1}{2} dt^{\prime} ] e^{- \Gamma_{0} t}   \nonumber\\
= - 2 \textit{E}_{X} r_{z}^{2} & \sin[a_{\texttt{k}}(t_{d}) t + b_{\texttt{k}}(t_{d}) \sin(2 \omega_{I} t + 2 \varphi) + \phi(t_{d})] e^{- \Gamma_{0} t}    \nonumber\\
= - 2 \textit{E}_{X} r_{z}^{2} & \lbrace\sin[a_{\texttt{k}}(t_{d}) t + \phi(t_{d})] \cdot \cos[b_{\texttt{k}}(t_{d}) \sin(2 \omega_{I} t + 2 \varphi)] \nonumber\\ 
+ \cos[& a_{\texttt{k}}(t_{d}) t + \phi(t_{d})] \cdot \sin[b_{\texttt{k}}(t_{d}) \sin(2 \omega_{I} t + 2 \varphi)]\rbrace e^{- \Gamma_{0} t},
\end{align}
where $a_{\texttt{k}}(t_{d}) = \epsilon_{c}(\texttt{k})  - \epsilon_{g} + \textit{A}_{I0}^{2} f_{I}^{2}(t_{d})/(4 m_{\texttt{k}}^{*})$, $b_{\texttt{k}}(t_{d})  = \textit{A}_{I0}^{2} f_{I}^{2}(t_{d})/(8\omega_{I} m_{\texttt{k}}^{*}) $, with effective mass $m_{\texttt{k}}^{*} = 1 / \nabla_{\texttt{k}}^{2} \epsilon_{c}(\texttt{k})$ for lattice momentum $\texttt{k}$, and $\phi(t_{d}) = -b_{\texttt{k}}(t_{d}) \sin(2 \varphi)$.

Utilizing Jacobi-Anger expansion formulas, Eq. (A4) can be further deduced to be 
\begin{widetext}
\begin{align}
\mu_{\texttt{k}}(t, &  t_{d}) = -2 E_{X} r_{z}^{2} \cdot J_{0}[b_{\texttt{k}}(t_{d})] \sin [a_{\texttt{k}}(t_{d}) t+\phi(t_{d})] e^{-\Gamma_{0} t} \nonumber\\
-2 & E_{X}  r_{z}^{2} \sum_{m=1}^{+\infty} J_{2 m}[b_{\texttt{k}}(t_{d})] \cdot \sin \left[ \left(a_{\texttt{k}}(t_{d})+4 m \omega_{I}\right) t+(\phi(t_{d}) + 4 m \varphi)\right] e^{-\Gamma_{0} t} \nonumber\\
-2 & E_{X} r_{z}^{2} \sum_{m=1}^{+\infty} J_{2 m}[b_{\texttt{k}}(t_{d})] \cdot \sin  \left[ \left(a_{\texttt{k}}(t_{d})-4 m \omega_{I}\right) t+(\phi(t_{d})-4 m \varphi)\right] e^{-\Gamma_{0} t} \nonumber\\
-2 & E_{X} r_{z}^{2} \sum_{m=1}^{+\infty} J_{2 m-1}[b_{\texttt{k}}(t_{d})] \cdot \sin  \lbrace\left[ (4 m-2) \omega_{I}+a_{\texttt{k}}(t_{d})\right] t+[(4 m-2) \varphi+\phi(t_{d})]\rbrace e^{-\Gamma_{0} t} \nonumber\\
-2 & E_{X} r_{z}^{2} \sum_{m=1}^{+\infty} J_{2 m-1}[b_{\texttt{k}}(t_{d})] \cdot \sin \lbrace\left[ (4 m-2) \omega_{I}-a_{\texttt{k}}(t_{d})\right] t+(4 m-2) \varphi-\phi(t_{d}) \rbrace e^{-\Gamma_{0} t},
\end{align}
\end{widetext}
where $J_{n}(x)$ is the $n$th-order Bessel function.

The response intensity is evaluated by
\begin{widetext}
\begin{align}
&S_{\texttt{k}}(\omega, t_{d})  \propto \operatorname{Im}[\tilde{\mu}_{\texttt{k}}(\omega, t_{d})] =  \operatorname{Im}[\int_{0}^{\infty} \mu_{\texttt{k}}(t,t_{d}) e^{-i \omega t} d t]  \nonumber\\
& = J_{0}[b_{\texttt{k}}(t_{d})] \cdot \lbrace L[\omega,a_{\texttt{k}}(t_{d})] \cdot \cos (\phi(t_{d})) +  F(\omega,a_{\texttt{k}}(t_{d})) \cdot \sin (\phi(t_{d})) \rbrace\nonumber\\
&+\sum_{m=1}^{+\infty} J_{2 m-1}[b_{\texttt{k}}(t_{d})] \cdot L\left[\omega,a_{\texttt{k}}(t_{d})+(4 m-2) \omega_{I}\right] \cdot \cos [\phi(t_{d})+(4 m-2) \varphi]\nonumber\\
&+\sum_{m=1}^{+\infty} J_{2 m-1}[b_{\texttt{k}}(t_{d})] \cdot F\left[\omega,a_{\texttt{k}}(t_{d})+(4 m-2) \omega_{I}\right] \cdot \sin [\phi(t_{d})+(4 m-2) \varphi]\nonumber\\
&-\sum_{m=1}^{+\infty} J_{2 m-1}[b_{\texttt{k}}(t_{d})] \cdot L\left[\omega,a_{\texttt{k}}(t_{d})-(4 m-2) \omega_{I}\right] \cdot \cos [\phi(t_{d})-(4 m-2) \varphi]\nonumber\\
&-\sum_{m=1}^{+\infty} J_{2 m-1}[b_{\texttt{k}}(t_{d})] \cdot F\left[\omega,a_{\texttt{k}}(t_{d})-(4 m-2) \omega_{I}\right] \cdot \sin [\phi(t_{d})-(4 m-2) \varphi]\nonumber\\
&+\sum_{m=1}^{+\infty} J_{2 m}[b_{\texttt{k}}(t_{d})] \cdot L\left[\omega,a_{\texttt{k}}(t_{d})+4 m \omega_{I}\right] \cdot \cos [\phi(t_{d})+4 m \varphi]\nonumber\\
&+\sum_{m=1}^{+\infty} J_{2 m}[b_{\texttt{k}}(t_{d})] \cdot F\left[\omega,a_{\texttt{k}}(t_{d})+4 m \omega_{I}\right] \cdot \sin [\phi(t_{d})+4 m \varphi]\nonumber\\
&+\sum_{m=1}^{+\infty} J_{2 m}[b_{\texttt{k}}(t_{d})] \cdot L\left[\omega,a_{\texttt{k}}(t_{d})-4 m \omega_{I}\right] \cdot \cos [\phi(t_{d})-4 m \varphi]\nonumber\\
&+\sum_{m=1}^{+\infty} J_{2 m}[b_{\texttt{k}}(t_{d})] \cdot F\left[\omega,a_{\texttt{k}}(t_{d})-4 m \omega_{I}\right] \cdot \sin [\phi(t_{d})-4 m \varphi],
\end{align}
\end{widetext}
where $L(\omega,x) = \dfrac{\Gamma_{0}}{\Gamma_{0}^{2} + (\omega - x)^{2}}$ and $F(\omega,x) = \dfrac{\omega - x}{\Gamma_{0}^{2} + (\omega - x)^{2}}$ are Lorentzian and Fano line shapes centered at $x$, respectively.

When the IR laser intensity and wavelength are $1 \times 10^{11}$ W/cm$^{2}$ and $3000$ nm, respectively, one can obtain $b_{\texttt{k}_{\Gamma}}(t_{d}=0) = \textit{A}_{I0}^{2}/(8\omega_{I} m^{*}_{\texttt{k}_{\Gamma}}) = -0.11$ and $b_{\texttt{k}_{\textsc{M}}}(t_{d}=0) = \textit{A}_{I0}^{2}/(8\omega_{I} m^{*}_{\texttt{k}_{\textsc{M}}}) = 0.33$.
Namely, $ |b_{\texttt{k}}(t_{d})| \leq 0.33$ for both the $\Gamma$ and $\textsc{M}$ points, and we can consider $J_{n \geq 2}[b_{\texttt{k}}(t_{d})] \approx 0$.
In addition, because of $ |b_{\texttt{k}}(t_{d})| \leq 0.33$, we can adopt two approximations $ \cos(\phi(t_{d})) = \cos[-b_{\texttt{k}}(t_{d}) \sin(2 \varphi)] \approx J_{0}[b_{\texttt{k}}(t_{d})] \approx 1$ and $\sin(\phi(t_{d})) = \sin[-b_{\texttt{k}}(t_{d}) \sin(2 \varphi)] \approx 0$.

According to Eq. (A6), the response intensity is reduced to
\begin{widetext}
\begin{align}
 S_{\texttt{k}} (\omega, t_{d}) &  \simeq  J_{0}[b_{\texttt{k}}(t_{d})] L(\omega,a_{\texttt{k}}(t_{d})) \nonumber\\
&+ J_{1}[b_{\texttt{k}}(t_{d})] L\left[\omega,a_{\texttt{k}}(t_{d})+2 \omega_{I}\right] \cos (2 \omega_{I} t_{d}) + J_{1}[b_{\texttt{k}}(t_{d})] F\left[\omega,a_{\texttt{k}}(t_{d})+2 \omega_{I}\right] \sin (2 \omega_{I} t_{d}) \nonumber\\
&- J_{1}[b_{\texttt{k}}(t_{d})] L\left[\omega,a_{\texttt{k}}(t_{d})-2 \omega_{I}\right] \cos (2 \omega_{I} t_{d}) + J_{1}[b_{\texttt{k}}(t_{d})] F\left[\omega,a_{\texttt{k}}(t_{d})-2 \omega_{I}\right] \sin (2 \omega_{I} t_{d}).
\end{align}
\end{widetext}
The ATAS spectrum of the electron with $\texttt{k}$ is evaluated by 
\begin{align}
\Delta S_{\texttt{k}} (\omega,t_{d}) = S_{\texttt{k}}(\omega, t_{d}) - S^{X}_{\texttt{k}}(\omega).
\end{align}

\end{document}